\documentstyle[prd,aps,epsfig]{revtex}
\begin{document}
\title{New Higgs Couplings at Tevatron \footnote{Contribution to
	the XIX International Symposium on Lepton and Photon Interactions 
	at High Energies, Stanford University, August 9--14, 1999.}}
\author{M.\ C.\ Gonzalez--Garcia $^{1}$, 
        S.\ M.\ Lietti $^2$ and 
        S.\ F.\ Novaes $^3$}
\address{
$^1$ Instituto de F\'{\i}sica Corpuscular IFIC/CSIC,
     Departament de F\'{\i}sica Te\`orica\\
     Universitat de Val\`encia, 46100 Burjassot, Val\`encia, Spain\\
$^2$ Theoretical Physics Group\\
     Ernest Orlando Lawrence Berkeley National Laboratory\\
     University of California, Berkeley, California 94720, USA\\
$^3$ Instituto de F\'{\i}sica Te\'orica, 
     Universidade  Estadual Paulista, \\  
     Rua Pamplona 145, 01405--900, S\~ao Paulo, Brazil.}
\maketitle
\begin{abstract}
We investigate the potentiality of CERN LEP and Fermilab Tevatron
colliders to establish bounds on new couplings involving the
bosonic sector of the standard model. A combined exclusion plot
for the coefficients of different dimension--6 anomalous
operators is presented. We also discuss the sensitivity that can
be achieved at the upgraded Tevatron.
\end{abstract}

%###############################################################################
\section*{Effective Lagrangians for Higgs Interactions} 

The effective Lagrangian is a model--independent approach to
describe new physics that is expected to manifest itself  at an
energy scale $\Lambda$, larger than the scale where the
experiments are performed. The effective Lagrangian can be
constructed out of higher dimensional operators and depends only
on the particle content of the low energy theory. We consider
here the possibility of having a light Higgs boson that should be
contained in these operators. Hence, we assume a linearly realized
\cite{linear,hisz} $SU_L(2) \times U_Y(1)$ invariant effective
Lagrangian to describe the bosonic sector of the SM, keeping the
fermionic sector unchanged. 

There are eleven dimension--6 operators involving the gauge
bosons and the Higgs scalar field which respect local $SU_L(2)
\times U_Y(1)$ and  $C$ and $P$ symmetries \cite{linear}.  Six of
these operators either affect only the Higgs self--interactions
or contribute to the gauge boson two--point functions at tree
level and are severely constrained from low energy physics below
the present sensitivity of high energy experiments \cite{hisz}.
From the remaining five ``blind'' operators, four affect the
Higgs couplings and can be written as
\cite{linear,hisz},
\begin{equation}
{\cal L}_{\mbox{eff}} = 
\frac{1}{\Lambda^2} \Bigl[
 f_W (D_{\mu} \Phi)^{\dagger} \hat{W}^{\mu \nu} (D_{\nu} \Phi) 
 + f_B (D_{\mu} \Phi)^{\dagger} \hat{B}^{\mu \nu} (D_{\nu} \Phi) 
+ f_{WW} \Phi^{\dagger} \hat{W}_{\mu \nu} \hat{W}^{\mu \nu} \Phi
+ f_{BB} \Phi^{\dagger} \hat{B}_{\mu \nu} \hat{B}^{\mu \nu} \Phi 
  \Bigr] 
\label{lagrangian}
\end{equation}
where $\Phi$ is the Higgs field doublet,   $\hat{B}_{\mu\nu} = i
(g'/2) B_{\mu \nu}$, and $\hat{W}_{\mu \nu} = i (g/2) \sigma^a
W^a_{\mu \nu}$ with $B_{\mu \nu}$ and $ W^a_{\mu \nu}$ being the
field strength tensors of the $U(1)$ and $SU(2)$ gauge fields
respectively. 

Anomalous $H\gamma\gamma$, $HZ\gamma$, and $HZZ$ and $HWW$
couplings are generated by (\ref{lagrangian}), which modify the
Higgs boson production and decay \cite{hagiwara2}. In the unitary
gauge they are given by 
\begin{eqnarray}
{\cal L}_{\mbox{eff}}^{\mbox{H}} &=& 
g_{H \gamma \gamma} H A_{\mu \nu} A^{\mu \nu} + 
g^{(1)}_{H Z \gamma} A_{\mu \nu} Z^{\mu} \partial^{\nu} H 
+ g^{(2)}_{H Z \gamma} H A_{\mu \nu} Z^{\mu \nu}
+ g^{(1)}_{H Z Z} Z_{\mu \nu} Z^{\mu} \partial^{\nu} H \nonumber \\
&+& g^{(2)}_{H Z Z} H Z_{\mu \nu} Z^{\mu \nu} +
g^{(2)}_{H W W} H W^+_{\mu \nu} W^{- \, \mu \nu} 
+ g^{(1)}_{H W W} \left (W^+_{\mu \nu} W^{- \, \mu} \partial^{\nu} H 
+ \mbox{h.c.} \right)\,
\label{H} 
\end{eqnarray}
where $A(Z)_{\mu \nu} = \partial_\mu A(Z)_\nu - \partial_\nu
A(Z)_\mu$. The effective couplings $g_{H \gamma \gamma}$,
$g^{(1,2)}_{H Z \gamma}$, and $g^{(1,2)}_{H Z Z}$  and
$g^{(1,2)}_{H WW}$ are related to the coefficients of the
operators appearing in (1) and can be found  elsewhere
\cite{hagiwara2}. Of special interest in our analysis is the
Higgs couplings to two photons which is given by 
\begin{equation}
g_{H \gamma \gamma} = 
- \left( \frac{g \sin^2\theta_W M_W}{2 \Lambda^2} \right)
                      (f_{BB} + f_{WW})  \; .
\label{g} 
\end{equation}

Equation (\ref{lagrangian}) also generates  new contributions to
the triple gauge boson vertex \cite{linear,hisz}. The operators
${\cal O}_{W}$  and ${\cal O}_{B}$ give rise to both anomalous
Higgs--gauge boson couplings and to new triple and quartic
self--couplings amongst the gauge bosons. On the other hand
${\cal O}_{WW}$ and ${\cal O}_{BB}$ only affect $HVV$ couplings
and cannot be constrained by the  study of anomalous trilinear
gauge boson couplings. 

\section*{Present Bounds from Searches at Tevatron Run I and Lep II}

Anomalous Higgs boson couplings have been studied in Higgs and
$Z^0$ boson decays \cite{hagiwara2}, in $e^+ e^-$
\cite{ee,our:lep2aaa,our:NLCWWA,our:NLCZZA} and in $p \bar{p}$
collisions
\cite{our:tevatronjj,our:tevatronmis,our:tevatron3a,our:combined}.
Let us first summarize the combined bounds on anomalous Higgs
boson interactions taking into account both Tevatron
\cite{d0jj,d0miss,cdf} and LEP \cite{opal:ggg} data on the
following signatures:
\begin{equation}
\begin{array}{lll}
~~~ \mbox{Process  } &  
~~~~~ \mbox{Anomalous Higgs Contribution}  &  ~~~ \mbox{Exp.\ Search}\\
&&\\
p\, \bar p  \rightarrow j \, j \, \gamma\, \gamma &  ~~~~~
p \bar{p} \to  W (Z) (\to j\, j) + H (\to \gamma \gamma) & ~~~~~
\mbox{D\O \protect\cite{d0jj}}
\\
p\, \bar p \rightarrow  \gamma \,\gamma \,+ \,\not \!\! E_T \nonumber &~~~~~
p \bar{p} \to  Z^0 (\to \nu \bar{\nu}) + H (\to \gamma \gamma) & ~~~~~
\mbox{D\O\protect\cite{d0miss}}  \\
& ~~~~~ p \bar{p} \to  W (\to [\ell] \nu) + H (\to \gamma \gamma) & ~~~~~
\nonumber \\ 
p\, \bar p  \rightarrow  \gamma\, \gamma\, \gamma & ~~~~~
p \bar p \rightarrow \gamma +H(\rightarrow \gamma \gamma) & ~~~~~
\mbox{CDF\protect\cite{cdf}} \\
e^+\, e^-  \rightarrow \gamma\, \gamma\, \gamma & ~~~~~
e^+ e^-  \rightarrow \gamma +H(\rightarrow \gamma \gamma) & ~~~~~
\mbox{OPAL  \protect\cite{opal:ggg} } 
\\
\label{proc}
\end{array}
\end{equation}
Events containing two photons plus missing energy,
additional photons or charged fermions represent a signature for
several theories involving physics beyond the SM and they have
been extensively searched for \cite{d0jj,d0miss,cdf,opal:ggg}. In
the framework of anomalous Higgs couplings presented before, they
can arise from the production of a Higgs boson which
subsequently decays in two photons [second column in
Eq.(\ref{proc})].  In the SM, the  decay width $H \to \gamma
\gamma$ is very small since it occurs just  at one--loop level
but  the existence of the new interactions (\ref{H})  can enhance
this width in a significant way. Recent analyses of these
signatures showed a good agreement with  the expectations from
the SM. Thus we can employ these negative experimental  results
to constrain new anomalous couplings in the bosonic sector of the
SM. 
\begin{figure}  [t!]
\centerline{\epsfig{file=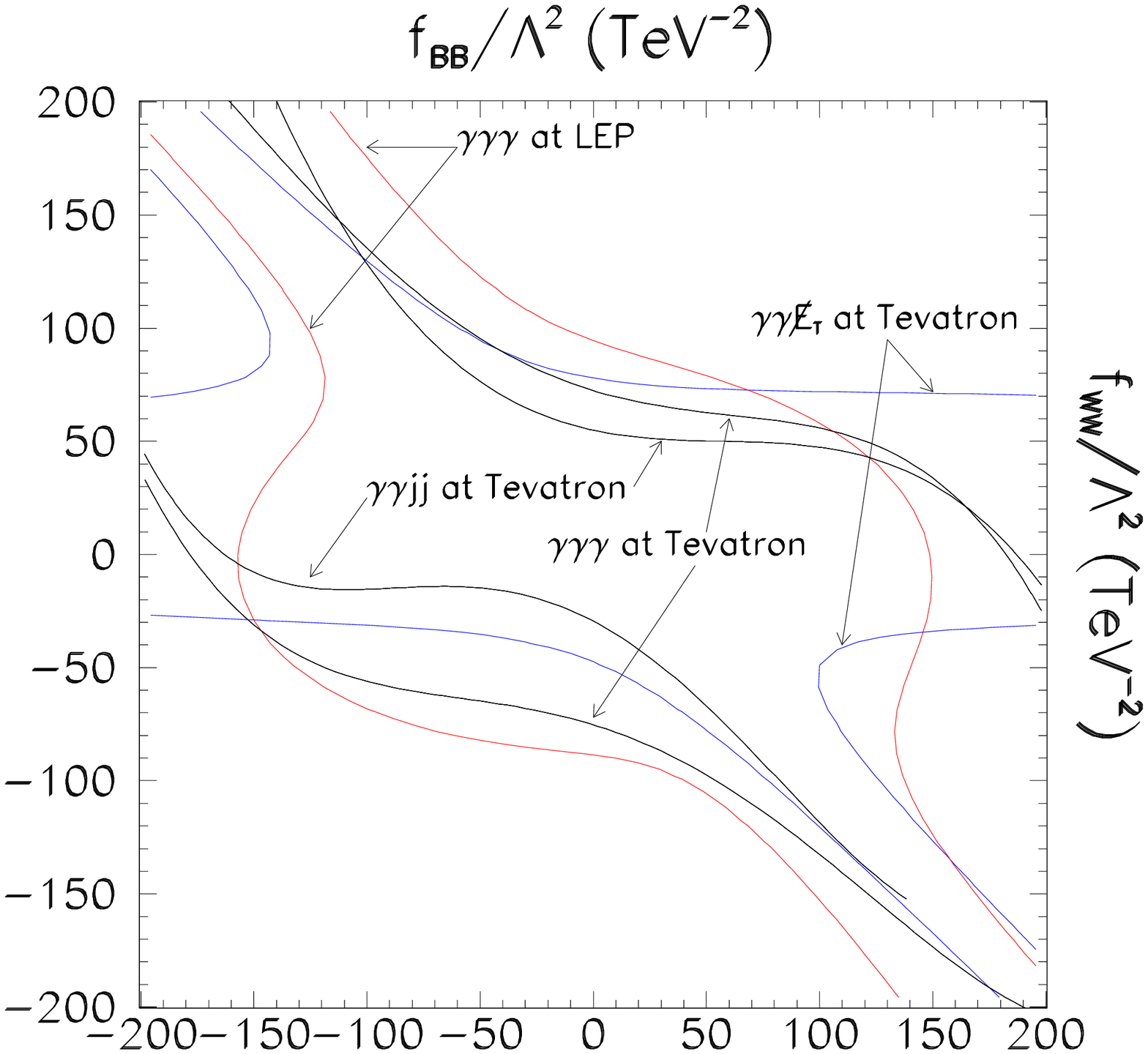,width=0.5\textwidth}
            \epsfig{file=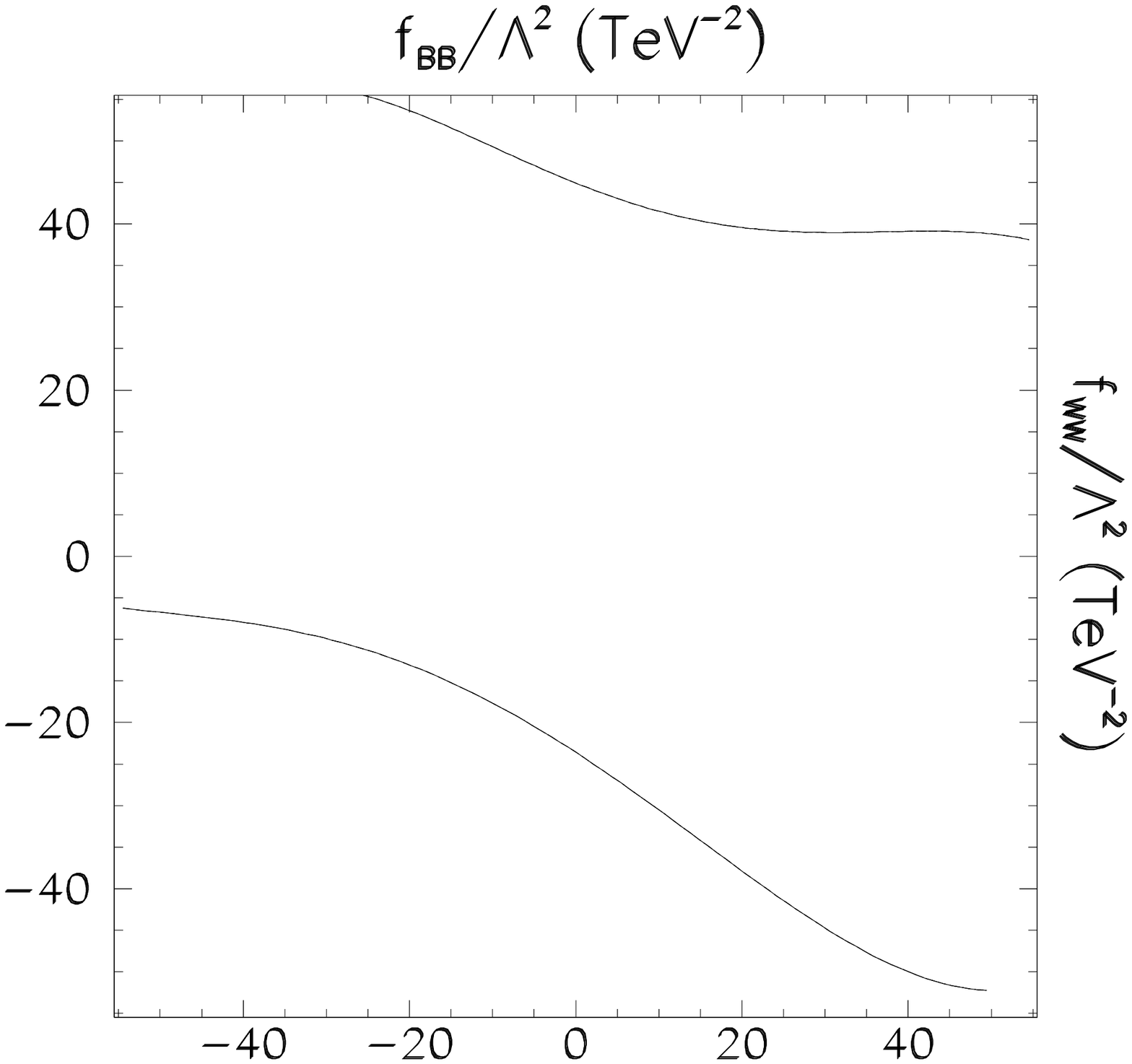,width=0.5\textwidth}}
\caption{{\bf (a)} Exclusion region outside the curves in the
$f_{BB} \times f_{WW}$ plane, in TeV$^{-2}$, based  on the D\O
~analysis \protect\cite{d0jj} of $\gamma\gamma j j$ production,
on  the D\O ~analysis \protect\cite{d0miss} of $\gamma\gamma \not
\!\! E_T$, on the CDF analysis \protect\cite{cdf} of
$\gamma\gamma\gamma$ production, and on the OPAL analysis
\protect\cite{opal:ggg} of $\gamma\gamma\gamma$ production,
always assuming $ M_H = 100$ GeV.  The curves show the 95\% CL
deviations from the SM total cross section. {\bf (b)} Same as
{(a)} for the combined analysis.}
\label{fig1}
\end{figure}

All processes listed in (\ref{proc}) have been the object of
direct experimental searches, and in our analysis we have closely
followed theses searches in order to make our study as realistic
as  possible. In this way we start by the process $p\bar{p} \to
W (Z) (\to j\, j)  + H (\to \gamma \gamma)$ \cite{our:tevatronjj}
to constrain the anomalous Higgs boson couplings described in
(\ref{H}). D\O ~Collaboration reported the results for the search
of high invariant--mass photon pairs in $p \bar{p} \to \gamma
\gamma j j$ events \cite{d0jj} at $\sqrt{s}=1.8$ TeV and $100$
pb$^{-1}$ of integrated luminosity where no event with
two--photon invariant mass  in the range $100 < M_{\gamma\gamma}
\lesssim 220$ were observed.  In our analysis, we applied the
same cuts of Ref.\ \cite{d0jj} and included the particle
identification and  trigger efficiencies. We have searched for
Higgs boson with mass in the range $100 < M_H \lesssim 220$,
since after the $WW(ZZ)$ threshold is reached the diphoton
branching ratio of Higgs is quite reduced.

For events containing two photons plus large missing transverse
energy ($\gamma\gamma \not \!\! E_T$) \cite{our:tevatronmis}  we
have used the results from  D\O ~collaborations \cite{d0miss}
which reported  that no event with two--photon invariant mass in
the range $100 < M_{\gamma\gamma} \lesssim 2 M_W$ was observed.
Anomalous Higgs couplings can give rise to this final state via
the contributions listed in the second column, third line of
Eq.(\ref{proc})  where in the second subprocess the charged
lepton ($[\ell] = e, \mu$) escapes undetected.  In order to
compare our predictions with the results of  D\O ~Collaboration
\cite{d0miss}, we have applied the same cuts of last article in
Ref.\ \cite{d0miss}.   After these cuts, we find that 80\% to 90\%
of the signal comes from associated Higgs--$Z^0$ production while
10\% to 20\% arrises from Higgs--$W$. We also include in our
analysis the particle identification and trigger efficiencies
which vary from 40\% to 70\% per photon. 

We have also analysed events with three photons in the final state
and compare our results  with the recent search reported by CDF
Collaboration \cite{cdf} for this signature. They looked for
$\gamma\gamma\gamma$ events requiring two photons in the
central region of the detector, with a minimum transverse energy
of 12 GeV, plus an additional photon with $E_T > 25$ GeV. The
photons were required to be separated by more than $15^\circ$.

Finally, for events containing three photons in the final 
state at electron--positron collisions \cite{our:lep2aaa},
we have used the recent OPAL results \cite{opal:ggg} where data
taken at several energy points in the range $\sqrt{s}=130$ --
$172$ GeV were combined. 

We have included in our calculations all SM (QCD plus
electroweak), and anomalous contributions that lead to these
final states. The SM one-loop contributions to the
$H\gamma\gamma$ and $H Z\gamma$ vertices were introduced through
the use of the  effective operators with the corresponding form
factors in the coupling.  Neither the narrow--width approximation
for the Higgs boson  contributions, nor the effective $W$ boson
approximation were employed. We consistently included the effect
of all interferences between the anomalous signature and the SM
background. For $p\,\bar p$ processes, we have used the  MRS (G)
\cite{mrs} set of proton structure functions with the scale
$Q^2=\hat{s}$.

\begin{figure} [t!]
\centerline{\epsfig{file=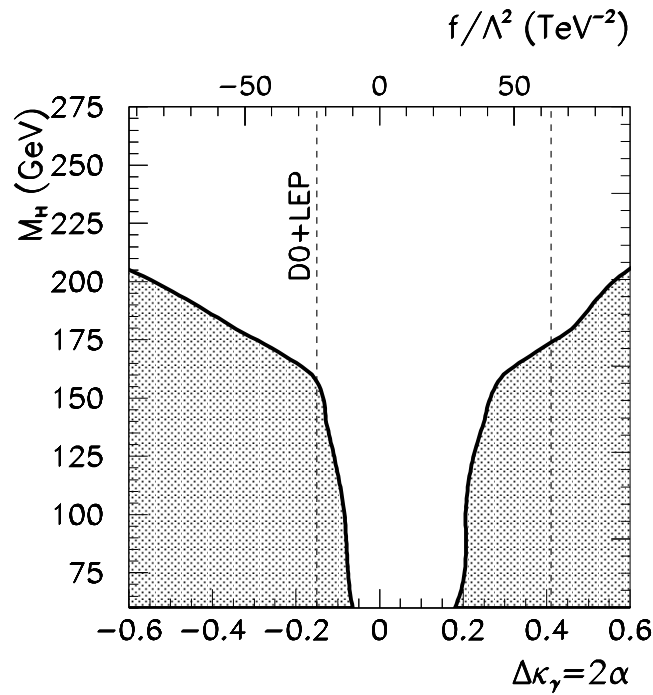,width=0.5\textwidth}}
\caption{Excluded region in the $f \times M_H$
plane from the combined analysis of the LEPII and Tevatron 
searches.}
\label{kappa}
\end{figure}

The coupling $H\gamma\gamma$ (\ref{g}) involves $f_{WW}$ and
$f_{BB}$ \cite{hagiwara2}.  In consequence, the anomalous
signature $f\bar f \gamma\gamma$ is only possible when those
couplings are not vanishing. The couplings $f_B$ and $f_W$, on
the other hand, affect the production mechanisms for the Higgs
boson. In Fig.\ \ref{fig1}.a we present our results for the
excluded  region in the $f_{WW}$, $f_{BB}$ plane from the
different channels  studied for $M_H = 100$ GeV, assuming that
these are the only non--vanishing couplings. Since the anomalous
contribution to  $H\gamma\gamma$ is zero for $f_{BB} = - f_{WW}$
(see Eq.\ \ref{g}), the bounds become very weak close to this
line, as is clearly shown in Fig.\ \ref{fig1}. In order to
establish these bounds, we imposed an upper limit on the number
of signal events  based on Poisson statistics. In the absence of
background  this implies $N_{\mbox{signal}} < 1 \,(3)$ at 64\%
(95\%) CL. In the presence of background events, we employed the
modified Poisson analysis. 

The results obtained from the analysis of the four reactions
(\ref{proc}) can be statistically  combined in order to constrain
the value of the coefficients $f_i$, $i=WW, BB,W, B$ of
(\ref{lagrangian}) \cite{our:combined}. We exhibit in Fig.\
\ref{fig1}.b the 95\% CL exclusion region in the plane $f_{BB}
\times f_{WW}$ obtained from combined results. 

In order to reduce the number of free parameters one can make the
assumption that all blind operators affecting the Higgs
interactions have a common coupling $f$, {\it i.e.} $f = f_W =
f_B = f_{WW} = f_{BB}$ \cite{hisz,hagiwara2}. 
In Fig \ref{kappa}, we present the combined limits for the coupling 
constant $f = f_{BB}= f_{WW}= f_{B} = f_{W}$ for Higgs boson masses 
in the range of $100 \leq M_H \leq 220$ GeV. 

\section*{Attainable Bounds at Future Tevatron Runs}

The effect of the anomalous operators becomes more evident with
the increase of energy, and therefore, higher sensitive to
smaller values of the anomalous coefficients can be achieved by
studying their contribution to different processes at the
upgraded Tevatron collider. We have considered the Tevatron Run
II upgrade, with 1 fb$^{-1}$, and the TeV33 upgrade with 10
fb$^{-1}$. For the reactions $p \bar{p} \to\gamma \gamma \not
\!\! E_T$ and $p \bar{p} \to \gamma \gamma j j$, we assumed the
same cuts and detection  efficiencies than at Tevatron Run I. For
the $\gamma\gamma\gamma$ final state we have studied the
improvement on the sensitivity to the anomalous coefficients by
implementing additional kinematical cuts \cite{our:tevatron3a}.
Best results are obtained for the following set of cuts:
$E_{T_{1}} > 40$ GeV, with $E_{T_{2,3}} > 12$ GeV, where we have
ordered the three photons according to their transverse energy,
{\it i.e.\/} $E_{T_{1}} > E_{T_{2}} > E_{T_{3}}$. We always
required the photons to be in the central region of the detector
($|\eta_{i}| < 1$) where there is sensitivity for electromagnetic
showering. In our estimate, we assumed the same detection
efficiency for photons as considered by CDF Collaboration
\cite{cdf} for the Run I.

\vskip 0.5cm
\begin{center}
\begin{tabular}{||c||c||c||c||c||}
\hline\hline
$M_H$(GeV) & \multicolumn{4}{c||}{$f/\Lambda^2$(TeV$^{-2}$)}  \\
\hline 
\hline
  & $p \bar{p} \to \gamma \gamma \gamma$  & $p \bar{p} 
  \to \gamma \gamma +  \!\! E_T $  &
  $p \bar{p} \to \gamma \gamma j j$  & Combined\\
\hline 
\hline
100 	& ($-$24, 24)  [$-$13, 15] 
	& ($-$16, 36)  [$-$9.4, 26] 
	& ($-$9.2, 22) [$-$3.3, 5.6] 
	& ($-$7.6, 19) [$-$3.0, 5.6]
\\
\hline
120 	& ($-$26, 26)  [$-$14, 14] 
	& ($-$20, 39)  [$-$15, 27]
	& ($-$8.6, 21) [$-$3.4, 5.9] 
	& ($-$7.4, 18) [$-$3.3, 5.9]
\\
\hline
140 	& ($-$30, 31)  [$-$15, 16] 
	& ($-$25, 44)  [$-$14, 30]
	& ($-$10, 23)  [$-$4.5, 8.9]
	& ($-$9.1, 20) [$-$4.0, 8.7]
\\
\hline
160 	& ($-$36, 38)  [$-$17, 19] 
	& ($-$29, 50)  [$-$14, 33]
	& ($-$11, 24)  [$-$6.0, 14]
	& ($-$9.9,22)  [$-$5.1, 13]
\\
\hline
180	& (---, ---)  [---, ---] 
	& ($-$63, 72) [$-$46, 53]
	& ($-$26, 34) [$-$16, 24]
	& ($-$24, 33) [$-$16, 24]  \\
\hline
200 	& (---, ---)  [---, ---]   
	& ($-$87, 90) [$-$50, 53]
	& ($-$33, 40) [$-$17, 23]
	& ($-$32, 39) [$-$17, 23] \\
\hline
220 	& (---, ---)  [---, ---]
	& (---, ---)  [---, ---]
	& ($-$42, 45) [$-$19, 26]
	& ($-$42, 45) [$-$19, 26] \\
\hline\hline
\end{tabular}

\vskip 0.3cm
TABLE I. ~ {\small 95\% CL allowed range for $f/\Lambda^2$,  from
$\gamma\gamma\gamma$, $\gamma\gamma +  \!\! E_T $,  $\gamma\gamma
j j $ production at Tevatron Run II [TeV33] assuming   all $f_i$
to be equal.  We denote by `---' limits worse than $|f|=100$
TeV$^-2$.}
\end{center}

\vskip 0.5cm
In Table I, we present the 95\% CL  limit on the
anomalous couplings for Tevatron Run II and TeV33  for each
individual process. All couplings are assumed equal  ($f =
f_{BB}= f_{WW}= f_{B} = f_{W}$) and the Higgs boson mass is
varied in the range $100 \leq M_H \leq 220$ GeV. Combination
of the results obtained from the analysis of the first three
reactions in Eq.(\ref{proc}) leads to the improved bounds given
in the last column of Table I. Comparing these
results with those in Fig.~\ref{kappa} we observe an improvement
of about a factor $\sim$ 2--3 ($\sim$ 4--6) for the combined
limits at Run II (TeV33).

As mentioned above for linearly realized effective  Lagrangians,
the modifications introduced in the Higgs  and in the vector
boson sector are  related to each other. In consequence the
bounds on the new Higgs couplings should also restrict the
anomalous gauge--boson self interactions.  Under the assumption
of equal coefficients for all anomalous Higgs operators, we can
relate the common Higgs boson anomalous coupling $f$ with the
conventional parametrization of the vertex $WWV$ ($V = Z^0$,
$\gamma$) \cite{hhpz},
\begin{equation}
\Delta \kappa_\gamma =\frac{M_W^2}{\Lambda^2}~ f =  
\frac{2 \cos^2 \theta_W }{1 - 2 \sin^2 \theta_W}~  
\Delta \kappa_Z= 2 \cos^2 \theta_W ~\Delta g^Z_1   
\label{trad} 
\end{equation}
A different set of parameters has been also used by the LEP
Collaborations in terms of  three independent couplings,
$\alpha_{B \Phi}$, $\alpha_{W \Phi}$, and $\alpha_{W}$, These
parameters are related to the parametrization of Ref.\
\cite{hhpz} through $\alpha_{B \Phi} \equiv \Delta \kappa_\gamma
- \Delta g^Z_1 \cos^2 \theta_W$, $\alpha_{W \Phi} \equiv \Delta
g^Z_1 \cos^2 \theta_W$, $\alpha_{W} \equiv \lambda_\gamma$.

The current experimental limit on these couplings  
from combined results on double gauge
boson production at Tevatron and LEP II \cite{vancouver}
is $-0.15\;<\;\Delta\kappa_\gamma\;=\;2\alpha \;<\;0.41$
at 95 \% CL. This limit is derived under the relations given in
Eq.\ (\ref{trad}) \cite{hisz}.

In Table II, we present the 95\% CL limit of the
anomalous coupling $\Delta\kappa_\gamma$ using the limits on
$f/\Lambda^2$ obtained  through the analysis of the processes
\ref{proc}.  We also present the expected bounds  that will be
reachable at the upgraded Tevatron. Our results show that the
present combined limit from the Higgs production analysis
obtained in this paper is comparable with the existing bound
from gauge boson production for $M_H \leq 170$ GeV.

\vskip 0.5cm
\begin{center}
\begin{tabular}{||c||c||}
\hline \hline 
& \\[-0.1cm]
Process  & ~ $\Delta\kappa_\gamma = 2 \,\alpha= 2 \,\alpha_{B \Phi} = 
2 \,\alpha_{W \Phi}$ ~ \\ 
&\\[-0.1cm]  
\hline \hline 
& \\[-0.1cm] 
Combined Tevatron Run I + LEP II &($-$0.084, 0.204) \\[0.2cm]
Combined Tevatron Run II & ($-$0.048, 0.122) \\ [0.2cm]
Combined Tevatron TeV33 & ($-$0.020, 0.036) \\
& \\[-0.1cm] 
\hline \hline 
\end{tabular}

\vskip 0.3cm
TABLE II. ~ {\small 95\% CL allowed range for the anomalous triple gauge boson 
couplings derived from the limits obtained for the anomalous Higgs boson 
coupling $f$ for $M_H= 100$ GeV.}
\end{center}

\vskip 0.5cm
Summarizing, we have estimated the limits on anomalous
dimension--six Higgs boson interactions that can be derived from
the investigation of three photon events at LEP2 and Tevatron and
diphoton plus missing transverse energy events or dijets at
Tevatron. Under the assumption that the coefficients of the four
``blind'' effective operators contributing to Higgs--vector boson
couplings are of the same magnitude, the study can give rise to a
significant indirect limit on anomalous $WWV$ couplings.  We have
also studied the expected improvement on the sensitivity  to
Higgs anomalous couplings at the  Fermilab Tevatron upgrades.

\vskip 0.4cm
\noindent
{\bf Acknowledgments:} 
This work was partially supported by Conselho Nacional de Desenvolvimento
Cient\'{\i}fico e Tecnol\'ogico (CNPq), by Funda\c{c}\~ao de
Amparo \`a Pesquisa do Estado de S\~ao Paulo (FAPESP), and by
Programa de Apoio a N\'ucleos de Excel\^encia (PRONEX).

%###############################################################################

%##############################################################################
\end{document}